\documentclass[prb,preprint,showpacs]{revtex4-1}
\usepackage{amsfonts}
\usepackage{amsmath}
\usepackage{amssymb}
\usepackage{lineno}
\usepackage{graphicx}%
\begin{document}

\linenumbers

\title{Ferromagnetic insulating state in tensile-strained LaCoO$_3$ thin films}

\author{Han Hsu,$^{1}$ Peter Blaha,$^{2}$ and Renata M. Wentzcovitch$^{1}$}

\affiliation{$^{1}$Department of Chemical Engineering and Materials
Science, University of Minnesota, Minneapolis, Minnesota, USA\\
$^2$Institute of Materials Chemistry, Vienna University of
Technology, A-1060 Vienna, Getreidemarkt 9/165-TC, Austria}

\date{\today}

\bigskip

\begin{abstract}

With local density approximation $+$ Hubbard $U$ (LDA+$U$)
calculations, we show that the ferromagnetic (FM) insulating state
observed in tensile-strained LaCoO$_3$ epitaxial thin films is most
likely a mixture of low-spin (LS) and high-spin (HS) Co, namely, a
HS/LS mixture state. Compared with other FM states, including the
intermediate-spin (IS) state (\textit{metallic} within LDA+$U$),
which consists of IS Co only, and the insulating IS/LS mixture
state, the HS/LS state is the most favorable one. The FM order in
HS/LS state is stabilized via the superexchange interactions between
adjacent LS and HS Co. We also show that Co spin state can be
identified by measuring the electric field gradient (EFG) at Co
nucleus via nuclear magnetic resonance (NMR) spectroscopy.

\end{abstract}


\maketitle

Perovskite-structure oxides have been proven a fertile area in
condensed matter physics. They exhibit amazing properties, including
ferroelectricity, ferromagnetism, colossal magnetoresistance (CMR),
and multiferroics (simultaneous ferroelectricity and
ferromagnetism), as a consequence of their spin, lattice, charge,
and orbital degree of freedom. Advances in thin-film growth
techniques have even brought more promising potentials for their
future application, as their properties can be engineered via
epitaxial strains. A few examples include strontium titanate
(SrTiO$_3$), ferroelectric in tensile-strained thin film while
paraelectric in bulk, \cite{ferro_STO} lanthanum titanate
(LaTiO$_3$), conducting in compressive-strained thin film while
insulating in bulk, \cite{ferro_LTO} and Europium titanate
(EuTiO$_3$), in which multiferroics induced by tensile strains has
been observed. \cite{multiferroic_EuTiO3} As to lanthanum cobaltite
(LaCoO$_3$), a diamagnetic insulator in bulk at low temperatures
($T<35$ K), a ferromagnetic (FM) \textit{insulating} state has been
observed in tensile-strained thin films, e.g. LaCoO$_3$ grown on
SrTiO$_3$ or (LaAlO$_3$)$_{0.3}$(Sr$_2$AlTaO$_6$)$_{0.7}$, at $T<85$
K, \cite{Fuchs 2007, Fuchs 2008, Pinta 2008, Fuchs 2009, Freeland
2008, Park 2009, Herklotz 2009, Mehta 2009, Merz 2010, Posadas 2011}
while the ferromagnetism induced by compressive strains, e.g.
LaCoO$_3$ grown on LaAlO$_3$, is not conclusive. \cite{Fuchs 2008,
Freeland 2008, Park 2009, Mehta 2011} Two questions arise
immediately: (1) Given that there are six $3d$ electrons in
Co$^{3+}$, which can thus have a total electron spin $S=0$, $1$, or
$2$, refereed to as low-spin (LS), intermediate-spin (IS), and
high-spin (HS) state, respectively, what is the spin state of Co in
FM LaCoO$_3$ thin films, in contrast to the LS Co in diamagnetic
bulk? (2) What leads to the formation of FM order in LaCoO$_3$ thin
films? After all, FM \textit{insulators} are rarely seen. So far,
all first-principles calculations have only found FM
\textit{metallic} LaCoO$_3$ thin films with all Co ions in IS state,
\cite{Posadas 2011,Gupta 2008, Rondinelli 2009} a prediction clearly
inconsistent with transport measurements. \cite{Freeland 2008}

\bigskip

While finite Co spin induced by tensile strains in LaCoO$_3$ thin
films has just started attracting attention, finite Co spin induced
by thermal excitation in bulk LaCoO$_3$ has been a highly
controversial issue for decades. \cite{Goodenough's review,Rao's
Reivew} With LS Co at $T<35$ K, bulk LaCoO$_3$ becomes a
paramagnetic insulator with finite Co spin at about $90$ K. Such a
spin-state crossover in the temperature range of $35$--$90$ K was
first suggested to be a LS-HS crossover \cite{Goodenough 1967,
config fluctuation 1, config fluctuation 2} but was later suggested
to be LS-IS based on a local density approximation + Hubbard $U$
(LDA+$U$) calculation. \cite{LS-to-IS LDA+U by Korotin} Since then,
both scenarios have received supports from various experimental and
theoretical works, but a consensus is not yet achieved (see
Ref.~\onlinecite{LaCoO3_EFG_Hsu} for a brief review). A study
regarding LaCoO$_3$ thin films may also help understanding LaCoO$_3$
bulk from a different perspective. In this paper, we investigate the
Co spin state in tensile-strained thin films and the formation of FM
order via a series of LDA$+U$ calculations. While LDA$+U$ has been
frequently used to study cobaltites and Co spin state, the choice of
Hubbard $U$ can be an issue. It has been shown that under the same
lattice parameter, the Hubbard $U$ affects the total energy and the
determination of ground state. \cite{Rondinelli 2009} A well
justified Hubbard $U$ determined by first principles would thus be
necessary for finding out the the actual ground state. In this
paper, we compute the Hubbard $U$ parameters of Co in all spin
states self-consistently with a linear response approach.
\cite{Cococcioni_LDA+U, Campo10,Ferric} This method has successfully
found the ground state of iron-bearing magnesium silicate
(MgSiO$_3$) perovskite at a wide range of pressure. \cite{Ferric}
Both the plane-wave pseudopotential (PWPP) method \cite{pseudo}
implemented in {\sc Quantum ESPRESSO} codes \cite{PWscf} and the
augmented plane wave + local orbitals (APW+lo) method \cite{APW+lo}
implemented in WIEN2k codes \cite{WIEN2k} are used. As shall be
pointed out later, the orbital occupancies of Co in thin films are
different from those in bulk, due to their different symmetries. We
therefore compute the electric field gradient (EFG) tensor at Co
nucleus, $V_{zz}$, with WIEN2k, to see whether the Co spin state in
thin films can be identified via EFG, as demonstrated in bulk.
\cite{LaCoO3_EFG_Hsu}

\bigskip

\begin{figure}[pt]
\begin{center}
\includegraphics[
]{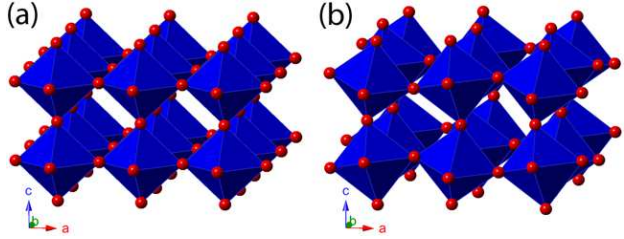}
\end{center}
\caption{(color online). Possible atomic structures of LaCoO$_3$
thin film (La is not shown) subject to constrained in-plane lattice
parameters. (a) Cube on cube, no CoO$_6$ octahedral rotation; (b)
full octahedral rotation degree of freedom.} \label{octa_rot}
\end{figure}

\begin{figure}[pt]
\begin{center}
\includegraphics[
]{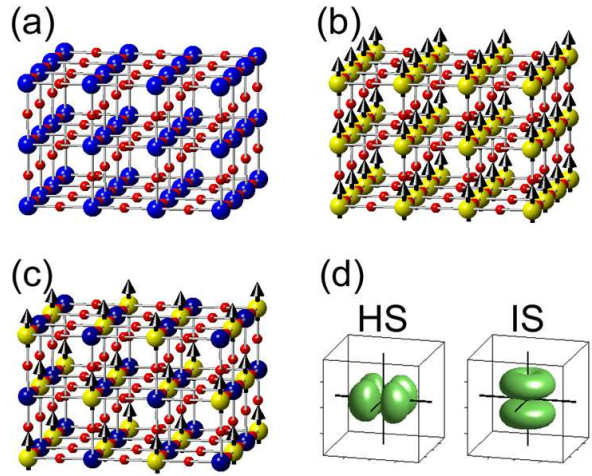}
\end{center}
\caption{(color online). (a)-(c) Possible magnetic configurations in
LaCoO$_3$ epitaxial thin film (La is not shown). The arrows denote
for nonzero magnetic moments, either IS or HS. (a) LS state; (b) HS
or IS state in FM order; (c) HS/LS or IS/LS mixture state in FM
order. (d) The $3d$ orbitals occupied by the spin-down electrons of
HS and IS Co.} \label{mag_config}
\end{figure}

The pseudocubic lattice parameter of bulk LaCoO$_3$
($R\overline{3}c$ symmetry) is about 3.81 {\AA} at $T \sim 5$
K.\cite{Radaelli and Cheong, zero point} To model tensile-strained
LaCoO$_3$ thin films via bulk calculations, we constrain the
in-plane pseudocubic lattice parameters $a_{\rm pc}$ and $b_{\rm
pc}$ of the hypothetical bulk to $3.899$ {\AA} (the lattice constant
of cubic SrTiO$_3$ at low temperatures), \cite{SrTiO3_abc} set
$\alpha=\beta=\gamma=90^{\circ}$, and optimize the out-of-plane
pseudocubic lattice parameter ($c_{\rm pc}$). Due to the lack of
accurate information regarding CoO$_6$ octahedral rotation in
thin-film LaCoO$_3$ in the low-temperature FM phase, we consider two
extreme cases shown in Fig.~\ref{octa_rot}: (a) cube-on-cube,
namely, no CoO$_6$ octahedral rotation degree of freedom, and (b)
full CoO$_6$ rotation degree of freedom subject to the
above-mentioned constraints. We also consider several magnetic
configurations shown in Fig.~\ref{mag_config}: (a) all Co ions in LS
state, (b) all Co ions have the same magnetic moment aligned in FM
order, and (c) a mixture state with LS Co surrounded by magnetic Co
(and vice versa) aligned in FM order. The configuration shown in
Fig. 2(c) is a legitimate postulate, as the observed magnetization
in LaCoO$_3$ thin films rarely exceeds 0.85 $\mu_B$/Co. \cite{Fuchs
2007, Fuchs 2008, Freeland 2008, Herklotz 2009, Park 2009, Mehta
2009, Merz 2010, Posadas 2011} For the configuration in
Fig.~\ref{mag_config}(b), a convergent wave function for HS state
cannot be obtained; only IS state can be found. For the
configuration in Fig.~\ref{mag_config}(c), both HS/LS and IS/LS
mixture state can be obtained. The self-consistent Hubbard $U$
parameter ($U_{\rm sc}$) of LS and IS Co are $7.0$ eV, while the
HS/LS state has $U_{\rm sc}^{\rm (HS)}=5.4$ and $U_{\rm sc}^{\rm
(LS)}=7.2$ eV. \cite{explain U} The dependence of $U_{\rm sc}$ on
$c_{\rm pc}$ is negligible. To demonstrate how the choice of Hubbard
$U$ can affect the determination of ground state, we also present
the result obtained using a constant $U=7$ eV for all Co in our PWPP
calculations. In tensile-strained LaCoO$_3$ thin films, CoO$_6$
octahedra possess tetragonal symmetry, namely, longer Co--O distance
on the $xy$ plane, regardless of CoO$_6$ rotation, as will be
discussed in the next paragraph. In tetragonal symmetry ($D_{4h}$),
the spin-down electron of HS Co occupies $d_{xy}$ orbital, and the
spin-down electrons of IS Co occupy $d_{xz}$ and $d_{yz}$ orbitals,
as shown in Fig.~\ref{mag_config}(d). Such orbital occupancies are
very different from those in bulk LaCoO$_3$, in which CoO$_6$
octahedra have trigonal symmetry ($D_{3d}$), with the [111]
direction being the high-symmetry axis. In bulk LaCoO$_3$, the
spin-down electron of HS Co occupies the $d_{z^2}$-like orbital
oriented along the [111] direction, and the IS Co spin-down
electrons occupy the doublet with 3-fold rotation symmetry about the
[111] direction. \cite{LaCoO3_EFG_Hsu}

\bigskip

The optimized out-of-plane pseudocubic lattice parameter ($c_{\rm
pc}$) of each FM state and associated relative energy ($\Delta E$)
and band gap ($E_{\rm gap}$) are listed in Tables I-II. Regardless
of CoO$_6$ rotation, the HS/LS mixture state (with $U_{\rm sc}$) is
the most stable FM state given by the PWPP method (Table I).
\cite{G-AFM} While the choice of $U=7.0$ eV makes HS/LS state less
favorable in PWPP calculations, APW+lo calculations still find HS/LS
the most stable FM state (Table II). Both PWPP and APW+lo methods
open an energy gap for HS/LS state, consistent with transport
measurements. \cite{Freeland 2008} Also, the presence of HS Co is
consistent with recent x-ray magnetic circular dichroism (XMCD) and
x-ray absorption spectroscopy (XAS) spectra. \cite{Mehta 2009, Merz
2010} In contrast, the IS state is never the most favorable FM
state, regardless of the computation method and CoO$_6$ rotation.
Its partially filled bands formed by partially occupied $d_{z^2}$
and $d_{x^2-y^2}$ orbitals in IS Co lead to a nonzero density of
state at the Fermi level. When the IS Co concentration is reduced to
50\%, an energy gap is opened in APW+lo calculations (Table II),
while PWPP calculations still give a conducting IS/LS state (Table
I). Such a difference is likely to result from the way that the
Hubbard $U$ is applied. In PWPP, the Hubbard $U$ is applied to the
projection of the total wave function onto Co $3d$ orbitals;
\cite{Cococcioni_LDA+U} in APW+lo, the Hubbard $U$ is directly
applied to the $3d$ orbitals within the muffin-tin radius of Co (1.9
bohr). Insulating or not, the IS/LS state is highly unlikely; its
energy is even higher than that of IS state. One more thing worthy
of mention is that the Co--O distance ($d_{\rm Co-O}$) and Co--O--Co
angle obtained in our calculation are different from those estimated
in Ref.~\onlinecite{Fuchs 2008}, where a constant $d_{\rm
Co-O}=1.93$ {\AA} is assumed (regardless of compressive or tensile
strains), and an Co--O--Co angle of $176^{\circ}$ is estimated for
LaCoO$_3$ grown on SrTiO$_3$. In our PWPP calculation with $U_{\rm
sc}$ and CoO$_6$ rotation, the HS/LS state has $d_{\rm
Co(HS)-O}=2.015$ and $1.872$ {\AA}, $d_{\rm Co(LS)-O}=1.922$ and
$1.886$ {\AA}, and Co--O--Co angles of $163.8^{\circ}$ and
$156.6^{\circ}$, on the ${xy}$ plane and along the $z$ axis,
respectively. Even for the IS state suggested by
Ref.~\onlinecite{Fuchs 2008}, we have $d_{\rm Co(IS)-O}=1.973$ and
$1.939$ {\AA}, and Co--O--Co angles of $162.3^{\circ}$ and
$154.9^{\circ}$, on the ${xy}$ plane and along the $z$ axis,
respectively.

\bigskip

Other than the total energy, structural properties can be a useful
criterion to determine which FM state favors tensile strains
($c_{\rm pc}/a_{\rm pc}<1$). Starting with the structures listed in
Table I with CoO$_6$ rotation, we perform full structural
optimization (at constrained volume) via variable cell-shape damped
molecular dynamics. \cite{VC-relax} All lattice parameters,
including $\alpha$, $\beta$, and $\gamma$, are optimized, so the
final structures only experience hydrostatic pressures. With
$\alpha$, $\beta$, and $\gamma$ slightly deviated from $90^{\circ}$,
the IS state has $c_{\rm pc}/a_{\rm pc}>1$, while all other states
remain $c_{\rm pc}/a_{\rm pc}<1$ (but IS/LS state still has a
$c_{\rm pc}/a_{\rm pc}$ larger than that of HS/LS state), as shown
in Table III. The larger $c_{\rm pc}/a_{\rm pc}$ ratio associated
with IS Co is a direct consequence of its occupied $d_{xz}$ and
$d_{yz}$ orbitals (by spin-down electrons), which elongate the Co--O
distance along the $z$-direction. In contrast, the fully optimized
HS/LS state has $c_{\rm pc}/a_{\rm pc}=0.969$, in great agreement
with $c_{\rm pc}/a_{\rm pc}=0.967$ observed in experiments.
\cite{Freeland 2008}

\bigskip

A significant part of cobalt-spin controversy arises from the
difficulty in directly measuring the total electron spin of Co. Such
difficulty, also appearing in other spin systems, can be resolved by
comparing the calculated and measured EFGs. \cite{QS MgFeSiO3 Hsu,
LaCoO3_EFG_Hsu, Ferric} So far, insulating FM state has been
observed in LaCoO$_3$ thin films with $a_{\rm pc}$ ranging from 3.84
to 3.90 {\AA}. \cite{Fuchs 2008, Herklotz 2009} In these thin films,
the magnetic Co concentration and Co--O distance may be different,
which can lead to slightly different EFG for Co in the same spin
state. To find out possible upper and lower limits of HS and IS Co
EFG, we compute them in two extreme cases: (1) thin films with
$a_{\rm pc}=3.899$ {\AA} and 50\% of magnetic Co, namely, the HS/LS
and IS/LS states listed in Tables I and II, and (2) single isolated
HS or IS Co in an array of LS Co in a fully relaxed structure with
$a_{\rm pc} \sim 3.81$ {\AA}, where the orbital occupancies of
isolated HS and IS Co are maintained in tetragonal symmetry
[Fig.~\ref{mag_config}(d)]. In these APW+lo calculations, IS Co does
not lead to a metallic state, in contrast to bulk LaCoO$_3$
($D_{3d}$ symmetry). \cite{LaCoO3_EFG_Hsu} Different choices of
Hubbard $U$ have been adopted as well (5 eV, 7 eV, and $U_{\rm
sc}$). The results of all these calculations show that the EFG
mainly depends on the spin state: $14.7 < V^{\rm
(HS)}_{zz}/(10^{21}$ V/m$^2)$ $<19.9$ and $-14.6 < V^{\rm
(IS)}_{zz}/(10^{21}$ V/m$^2)$ $<-8.0$. The quadrupole frequency,
$\nu_{Q}\equiv3eQ|V_{zz}|/2I\left( 2I-1\right) h$, can thus be
easily predicted, with $Q=0.42\times10^{-28}$ m$^2$ and $I=7/2$ for
$^{59}$Co nucleus. Based on the range of $V^{\rm (HS)}_{zz}$ and
$V^{\rm (IS)}_{zz}$, we conclude that in insulating LaCoO$_3$ thin
films, a measured $\nu_Q$ via nuclear magnetic resonance (NMR)
spectroscopy within $\sim 8.2 \pm 2.4$ MHz can be a strong evidence
for IS Co, and a measured $\nu_Q$ within $\sim 12.6 \pm 1.9$ MHz
should indicate HS Co.

\bigskip

\begin{figure}[pt]
\begin{center}
\includegraphics[
]{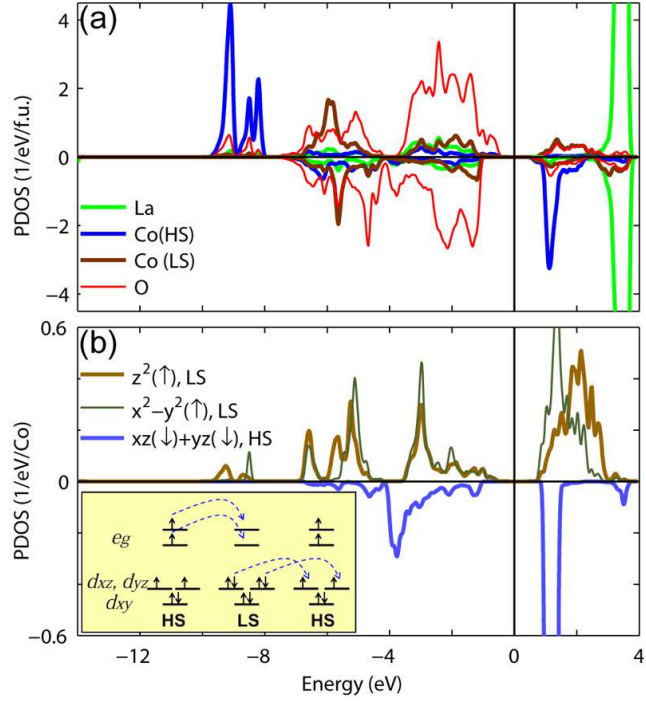}
\end{center}
\caption{(color online). Projected density of states of
ferromagnetic HS/LS state (no CoO$_6$ rotation, $U=7$ eV) onto (a)
each atomic site, and (b) some of the Co $3d$ orbitals. The inset in
(b) shows the electron transfer between HS and LS Co.} \label{PDOS}
\end{figure}

Analysis of electronic structures can help developing a physical
understanding for the FM order in HS/LS state, whose projected
density of states (PDOS) are shown in Fig.~\ref{PDOS}. The case with
$U=7$ eV and no CoO$_6$ rotation is presented, as the main features
in PDOS are not sensitive to the choice of $U$ and CoO$_6$ rotation.
Extracted from Fig.~\ref{PDOS}(a), both HS and LS Co have nonzero
magnetic moment: $2.97$ and $0.56$ $\mu_B$, respectively. The FM
order is established via the superexchange interaction between HS
and LS Co, as described by the Goodenough-Kanamori rule,
\cite{Goodenough's review, G-K rule 1, G-K rule 2, G-K rule 3} which
states that the superexchange interaction between two cations (with
or without a shared anion) is ferromagnetic if the electron transfer
is from a filled to a half-filled orbital or from a half-filled to
an empty orbital. Indeed, for the HS/LS state, electrons transfer
from the filled $d_{xz}$ and $d_{yz}$ orbitals of LS Co to the
half-filled $d_{xz}$ and $d_{yz}$ orbitals of HS Co via the oxygen
in between, and also from the half-filled $e_g$ ($d_{z^2}$ and
$d_{x^2-y^2}$) orbitals of HS Co to the empty $e_g$ orbitals of LS
Co, as depicted in the inset of Fig.~\ref{PDOS}(b). The PDOS shown
in Fig.~\ref{PDOS}(b) confirms this model: the finite spin-up $e_g$
electrons localized at the LS Co site (transferred from the HS Co
site) and the finite spin-down $d_{xz}$ and $d_{yz}$ electrons
localized at the HS Co site (transferred from the LS Co site). Such
electron transfers have been also described via a
\textit{configuration fluctuation} model, \cite{config fluctuation
1, config fluctuation 2} which further suggests that the interchange
of spin states (without net transfer of charge) led by electron
transfers stabilizes the FM order in this HS/LS states.

\bigskip

\begin{figure}[pt]
\begin{center}
\includegraphics[
]{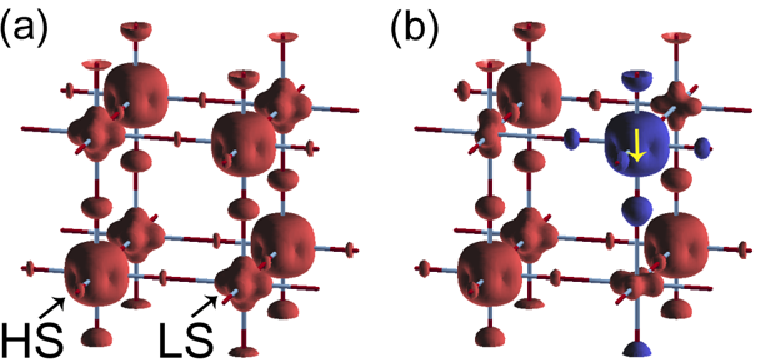}
\end{center}
\caption{(color online). Spin density, $s(\mathbf{r})$, of HS/LS
state (no CoO$_6$ rotation, $U=7$ eV) with (a) all HS Co magnetic
moments aligned, and (b) one HS Co magnetic moment flipped downward
(indicated by arrow). The isosurface values are $0.02$ (red) and
$-0.02$ (blue).} \label{spin}
\end{figure}

The above-mentioned superexchange interaction can be visualized via
electron spin density $s(\mathbf{r}) \equiv
\rho_{\uparrow}(\mathbf{r})-\rho_{\downarrow}(\mathbf{r})$, where
$\rho_{\uparrow}(\mathbf{r})$ and $\rho_{\downarrow}(\mathbf{r})$
are spin-up and spin-down electron density, respectively.
Figure~\ref{spin}(a) shows $s(\mathbf{r})$ corresponding to the
configuration with all HS Co magnetic moments aligned (same as the
configuration in Fig.~\ref{PDOS}). The nonzero magnetic moments
localized at LS Co sites (with $e_g$ character), aligned with the HS
Co magnetic moments, are consistent with the PDOS shown in
Fig.~\ref{PDOS}. When the magnetic moment of one HS Co in a 40-atom
supercell is flipped [Fig.~\ref{spin}(b)], the alignment of magnetic
moments is altered, and so is the condition that allows
configuration fluctuation [inset of Fig.~\ref{PDOS}(b)]. The spin
density at surrounding LS Co sites is thus significantly affected.
One flipped HS Co spin (in a 40-atom cell) increases the total
energy by 195 meV/supercell. Flipping one more HS Co spin, so the
total magnetization per supercell becomes zero, further increases
the total energy by 78 meV/supercell. With CoO$_6$ rotation, the
energy increases associated with one and two flipped HS Co spins are
96 and 34 meV/supercell, respectively. These results indicates that
the magnetic moment of HS Co in the HS/LS state should align at low
temperatures.

\bigskip

While our calculations have shown that HS/LS state is the most
favorable state among the ferromagnetic states being considered,
magnetic state in actual LaCoO$_3$ thin films can be more
complicated. The magnetization observed in experiments rarely
exceeds 0.85 $\mu_B$/Co,\cite{Fuchs 2007, Fuchs 2008, Freeland 2008,
Herklotz 2009, Park 2009, Mehta 2009, Merz 2010, Posadas 2011}
smaller than that of the HS/LS mixture (2 $\mu_B$/Co). Such a
magnetization suggests that the HS Co population should be smaller
than 50\%. In fact, XAS spectra combined with atomic multiplet
calculations have suggested that LaCoO$_3$ thin film on SrTiO$_3$
consists of about 64\% of LS Co and 36\% of HS Co. \cite{Merz 2010}
Given that the FM order is achieved via the superexchange
interaction within the HS-LS-HS Co configuration shown in
Fig.~\ref{mag_config}(c), one can thus expect ferromagnetic HS/LS
domain and nonmagnetic LS domain coexist in tensile strained thin
films, as observed using magnetic force microscopy (MFM). \cite{Park
2009} Also, since HS/LS state favors larger in-plane lattice
parameter, thin films with larger in-plane lattice parameters can be
expected to have larger HS/LS domain, and thus larger magnetization,
consistent with the increase of magnetization with lattice parameter
observed in experiment. \cite{Fuchs 2008}

\bigskip

In summary, we use LDA+$U$ calculations to show that the
ferromagnetic insulating state in tensile-strained LaCoO$_3$ thin
films is most likely a mixture of HS and LS Co. Among all the
ferromagnetic states studied in this paper (HS/LS, IS/LS, and IS),
the insulating HS/LS mixture state is the most favorable one,
energetically and structurally. Its FM order is established via the
superexchange interaction between LS and HS Co. We also show that
cobalt spin states in LaCoO$_3$ thin films could be identified via
NMR spectroscopy.

\bigskip

This work was primarily supported by the MRSEC Program of NSF grants
DMR-0212302 and DMR-0819885, and partially supported by EAR-081272
and EAR-1047629. P.B. was supported by the Austrian Science Fund
(SFB F41, "ViCoM"). Calculations were performed at the Minnesota
Supercomputing Institute (MSI). We thank C. Leighton for valuable
discussions.

\newpage

\begin{table}
\caption{Optimized out-of-plane pseudocubic lattice parameter
($c_{\rm pc}$) and associated relative energy ($\Delta E$) and
energy gap ($E_{\rm gap}$) of each FM state in tensile-strained
LaCoO$_3$ thin film (PWPP method).}
\begin{tabular}
[c]{cccccccc}\hline\hline & \multicolumn{3}{c}{No CoO$_{6}$
rotation} &  & \multicolumn{3}{c}{Full CoO$_{6}$
rotation}\\\cline{2-4}\cline{6-8} & $\ c_{\rm pc}$ (\AA ) \  & $\
\Delta E$ (eV/f.u.) \  & $\ E_{\rm gap}$ (eV) \  &   & $\ \ c_{\rm
pc}$ (\AA ) \  & $\ \Delta E$ (eV/f.u.) \  & $\ E_{\rm gap}$ (eV)
\ \\\cline{1-4}\cline{6-8}%
LS & $3.865$ & $0.35$ & $0.54$ &  & $3.660$ & $0.32$ & $1.24$\\\cline{1-4}%
\cline{6-8}%
IS & $3.785$ & $0.20$ & metal &  & $3.785$ & $0.19$ & metal\\\cline{1-4}%
\cline{6-8}%
IS/LS & $3.720$ & $0.35$ & metal &  & $3.680$ & $0.29$ & metal\\\cline{1-4}%
\cline{6-8}%
HS/LS ($U_{sc}$) & $3.685$ & $0.00$ & $0.92$ &  & $3.680$ & $0.00$ &
$0.90$\\\cline{1-4}\cline{6-8}%
HS/LS ($U=7$ eV) & $3.700$ & $0.29$ & $1.12$ &  & $3.695$ & $0.29$ &
$0.90$\\\hline\hline
\end{tabular}
\end{table}

\newpage

\begin{table}
\caption{Optimized $c_{\rm pc}$ and associated $\Delta E$ and
$E_{\rm gap}$ of each FM state (APW+lo method, with CoO$_6$
rotation).}

\begin{tabular}
[c]{cccc}\hline\hline & $\ \ c_{\rm pc}$ (\AA) & $\ \Delta E$
(eV/f.u.) & $\ E_{\rm gap}$ (eV) \\\hline LS & $\ 3.660$ \  & $\
0.37$ \ & $\ 1.72$ \
\\\hline IS & $3.741$ & $0.18$ & metal\\\hline IS/LS & $3.672$ &
$0.29$ & $0.59$\\\hline HS/LS ($U=7$ eV) & $3.686$ & $0$ &
$1.52$\\\hline\hline
\end{tabular}
\end{table}

\newpage

\begin{table}
\caption{Fully optimized pseudocubic lattice parameters of each FM
state at the volume as in Table I (with CoO$_6$ rotation).}

\begin{tabular}
[c]{cccc}\hline\hline & $\ a_{\rm pc}$, $b_{\rm pc}$ (\AA) & $\
c_{\rm pc}$ (\AA) & $\ c_{\rm pc}/a_{\rm pc}$ \ \\\hline IS & $\
3.848$ \  & $\ 3.893$ \ & $\ 1.012$ \ \\\hline IS/LS & $3.847$ &
$3.778$ & $0.982$\\\hline HS/LS ($U_{sc}$) & $3.863$ & $3.745$ &
$0.969$\\\hline HS/LS ($U=7$ eV) & $3.865$ & $3.757$ &
$0.972$\\\hline\hline
\end{tabular}
\end{table}

\newpage

\newpage

\newpage

\newpage


\begin{thebibliography}{9}

\bibitem{ferro_STO} J. H. Haeni \textit{et al}., Nature \textbf{430},
758 (2004).

\bibitem{ferro_LTO} F. J. Wong \textit{et al}., Phys. Rev. B \textbf{81},
161101(R) (2010).

\bibitem{multiferroic_EuTiO3} J. H. Lee \textit{et al}., Nature \textbf{466},
954 (2010).

\bibitem{Fuchs 2007} D. Fuchs \textit{et al}., Phys. Rev. B \textbf{75},
144402 (2007).

\bibitem{Fuchs 2008} D. Fuchs \textit{et al}., Phys. Rev. B \textbf{77},
014434 (2008).

\bibitem{Freeland 2008} J. W. Freeland \textit{et al}., Appl. Phys. Lett.
\textbf{93}, 212501 (2008).

\bibitem{Park 2009} S. Park \textit{et al}., Appl. Phys. Lett.
\textbf{95}, 072508 (2009).

\bibitem{Herklotz 2009} A. Herklotz \textit{et al}., Phys. Rev. B
\textbf{79}, 092409 (2009).

\bibitem{Mehta 2009} V. V. Mehta \textit{et al}., J. Appl. Phys.
\textbf{105}, 07E503 (2009).

\bibitem{Merz 2010} M. Merz \textit{et al}., Phys. Rev. B \textbf{82},
174416 (2010).

\bibitem{Posadas 2011} A. Posadas \textit{et al}., Appl. Phys. Lett.
\textbf{98}, 053104 (2011).

\bibitem{Pinta 2008} C. Pinta \textit{et al}., Phys. Rev. B \textbf{78},
174402 (2008).

\bibitem{Fuchs 2009} D. Fuchs \textit{et al}., Phys. Rev. B \textbf{79},
024424 (2009).

\bibitem{Mehta 2011} V. Mehta and Y. Suzuki, J. Appl. Phys. \textbf{109},
07D717 (2011).

\bibitem{Gupta 2008} K. Gupta and P. Mahadevan, Phys. Rev. B \textbf{79},
020406(R) (2009).

\bibitem{Rondinelli 2009} J. M. Rondinelli and N. A. Spaldin, Phys. Rev.
B \textbf{79}, 054409 (2009).

\bibitem{Goodenough's review} J. B. Goodenough, \textit{Localized
to Itinerant Electronic Transition in Perovskite Oxides} (Springer,
2001).

\bibitem{Rao's Reivew} C. N. R. Rao \textit{et al}., Top. Curr.
Chem. \textbf{234}, 1 (2004).

\bibitem{Goodenough 1967} P. M. Raccah and J. B. Goodenough, Phys.
Rev. \textbf{155}, 932 (1967).

\bibitem{config fluctuation 1} M. A. Se\~{n}ar\'{\i}s-Rodr\'{\i}guez
and J. B. Goodenough, J. Solid State Chem. \textbf{116}, 224 (1995).

\bibitem{config fluctuation 2} M. A. Se\~{n}ar\'{\i}s-Rodr\'{\i}guez
and J. B. Goodenough, J. Solid State Chem. \textbf{118}, 323 (1995).

\bibitem{LS-to-IS LDA+U by Korotin} M. A. Korotin \textit{et al}.,
Phys. Rev. B \textbf{54}, 5309 (1996).

\bibitem{LaCoO3_EFG_Hsu} Han Hsu \textit{et al}., Phys. Rev. B \textbf{82},
100406(R) (2010).

\bibitem{Cococcioni_LDA+U} M. Cococcioni and S. de Gironcoli, Phys. Rev.
B \textbf{71}, 035105 (2005).

\bibitem{Campo10} V. L. Campo Jr and M. Cococcioni, J. Phys.: Condens. Matter
\textbf{22}, 055602 (2010).

\bibitem{Ferric} Han Hsu \textit{et al}., Phys. Rev. Lett. \textbf{106},
118501 (2011).

\bibitem{pseudo} Pseudopotentials used in this paper have been reported
in H. Hsu \textit{et al}., Phys. Rev. B \textbf{79}, 125124 (2009).

\bibitem{PWscf} P. Giannozzi \textit{et al}., J. Phys.: Condens. Matter
\textbf{21}, 395502 (2009).

\bibitem{APW+lo} G. Madsen \textit{et al}., Phys. Rev. B \textbf{64},
195134 (2001).

\bibitem{WIEN2k} P. Blaha \textit{et al}., \textit{WIEN2k, An Augmented
Plane Wave Plus Local Orbitals Program for Calculating Crystal
Properties}, edited by K. Schwarz, Techn. Universit\"{a}t Wien,
Vienna (2001).

\bibitem{Radaelli and Cheong}P. G. Radaelli and S.-W. Cheong, Phys.
Rev. B \textbf{66}, 094408 (2002).

\bibitem{zero point} While LDA+$U$ gives a pseudocubic lattice parameter
of about $3.78$ {\AA} for bulk LaCoO$_3$ (see
Ref.~\onlinecite{pseudo}), inclusion of zero point motion energy can
improve this 1\% underestimate, and better agreement with
experiments can be achieved, as reviewed by R. M. Wentzcovitch
\textit{et al}., Rev. Min. Geochem. \textbf{71}, 59 (2010).

\bibitem{SrTiO3_abc} F. W. Lytle, J. Appl. Phys. \textbf{35}, 2212
(1964).

\bibitem{explain U} In this paper, $U_{\rm sc}$ is extracted from a series
of LDA+$U$ ground states associated with different trial $U$, as
detatiled in the supplemental material of Ref.~\onlinecite{Ferric}
(http://link.aps.org/supplemental/10.1103/PhysRevLett.106.118501).
In our earlier work on LS LaCoO$_3$ (Ref.~\onlinecite{pseudo}), the
Hubbard $U$ for LS Co ($\sim 8.3$ eV) was extracted from the LDA
ground state, as detailed in Ref.~\onlinecite{Cococcioni_LDA+U}.

\bibitem{G-AFM} In our calculations with constrained in-plane lattice
parameters ($a_{\rm pc}=b_{\rm pc}=3.899$ {\AA}), a HS state with
G-type antiferromagnetic (AFM) order can be obtained, and its energy
is lower than that of the HS/LS state. Given the neglegct of finite
thickness, the possible uncertainty of energy given by LDA+$U$
method, and the fact that AFM thin film is not observed in
experiments, we limit our discussions on the available FM states
(IS, IS/LS, and HS/LS).

\bibitem{VC-relax} R. M. Wentzcovitch \textit{et al}., Phys. Rev. Lett.
\textbf{70}, 3947 (1993).

\bibitem{QS MgFeSiO3 Hsu} Han Hsu \textit{et al}., Earth Planet. Sci.
Lett. \textbf{294}, 19 (2010).

\bibitem{G-K rule 1} J. B. Goodenough, Phys. Rev. \textbf{100}, 564
(1955).

\bibitem{G-K rule 2} J. B. Goodenough, J. Phys. Chem. Solids \textbf{6},
287 (1958).

\bibitem{G-K rule 3} J. Kanamori, J. Phys. Chem. Solids \textbf{10},
87 (1959).

\end{thebibliography}
\end{document}